\documentclass[12pt]{article}
\usepackage{amsfonts}
\usepackage{amsmath}
\usepackage{epsfig}

\begin{document}

\begin{flushright}
12/2012\\
\end{flushright}
\vspace{20mm}
\begin{center}
\large {\bf COLOR AND ISOSPIN WAVES FROM TETRAHEDRAL SHUBNIKOV GROUPS}\\
\mbox{ }\\
\normalsize
\vspace{1.0cm}
{\bf Bodo Lampe} \\              
\vspace{0.3cm}
II. Institut f\"ur theoretische Physik der Universit\"at Hamburg \\
Luruper Chaussee 149, 22761 Hamburg, Germany \\
\vspace{3.0cm}
{\bf Abstract}\\
\end{center} 
This letter supplements a recent 
article \cite{lamm} in which it was pointed out that the observed 
spectrum of quarks and leptons can arise as quasi-particle 
excitations in a discrete internal space. The paper concentrated 
on internal vibrational modes and it was only noted in the end that 
internal spin waves ('mignons') might do the same job. 
Here it will be shown how the mignon-mechanism works in detail.
In particular the Shubnikov group $A_4 + S ( S_4 - A_4)$ will be used 
to describe the spectrum, and the mignetic ground state is 
explicitly given. 



\newpage

\normalsize




In section 7 of ref. \cite{lamm} 
8 internal spins were considered, whose collective 
excitations ('mignons' or 'i-spin waves') were used to construct
quark and lepton states.
The scenario to start with 
is a spinor model in a (6+2)-dimensional 
spacetime which by some unknown compactification process
splits into a (3+1)-dimensional Minkowski space plus 
a (3+1)-dimensional internal space, in such a way 
that the internal space reappears as a finite and 
discrete internal 'crystal' on each point of the physical base 
space \cite{lamm}.


Generally, in $SO(d_1,d_2)$ the spinor dimensions viewed over complex space 
coincide with the case of the $(d_1+d_2)$-dimensional Euclidean space.
Therefore spinors in 6+2 dimensions can be considered as SO(8) spinors 
by a suitable Wick rotation. 
Note that the Lie algebras of SO(8) as well as SO(6,2) 
and SO(4,4) are extremely symmetric - the 
key word here is triality \cite{dixonbook} - and have an intimate connection 
to the nonassociative division algebra of octonions \cite{conway,kantor}. 
In fact, triality induces a bilinear multiplication on the 
8-dimensional representation space which is nothing else 
than the multiplication of octonions. 
This point will become important later, because octonion 
multiplication allows to connect the chiralities in the internal 
and in physical space. 
Furthermore, it is known that SO(8)-spinors can appear as 8-dimensional right-handed 
states $\mathbf{8_R}$ as well as 8-dimensional left-handed states $\mathbf{8_L}$
\cite{rossbook}. Both representations can be 
combined to a 16-dimensional Dirac spinor $\mathbf{8_L} + \mathbf{8_R}$ 
in a similar way as a Dirac spinor in 3+1 dimensions can be 
written as a sum of two Weyl spinors $\mathbf{2_L} + \mathbf{2_R}$. 

When going to $SO^{ph}(3,1)\times SO^{in}(3,1)$ the SO(6,2) spinor will 
split into a product of a Dirac spinor in internal space and 
a Dirac spinor on Minkowski space according to \cite{slansky} 
\begin{eqnarray}  
\mathbf{8_L} + \mathbf{8_R} = (\mathbf{2_L^{ph}} + \mathbf{2_R^{ph}}, \mathbf{2_L^{in}} + \mathbf{2_R^{in}}) 
\label{eq3uijj}
\end{eqnarray}
The corresponding field F has therefore 2 spinor indices
a and i both running from 1 to 4 and corresponding to a particle 
with Dirac properties both in physical and in internal space.

The next step is to assume that after the compactification process 
a copy of the internal space $I_x$ is fixed
to each point $x$ of physical space, so that internal Lorentz symmetry 
is broken and the induced i-spin structure  
can be analyzed as a nonrelativistic system of 
strongly correlated 3-dimensional i-spin vectors, 
the latter with an internal $SO^{in}(3)$ symmetry. 
This is not only supported by phenomenological 
observations (see below) but as a benefit the methods of solid state 
physics for the description of magnetic systems can be applied.

More in detail, the strongly correlated internal dynamics is described 
in second quantization language by 
creation and annihilation operators satisfying the 
canonical anti-commutation relations 
\begin{eqnarray} 
[c_{\alpha} (m), c_{\beta}^{\dagger} (m')]_- =\delta_{\alpha\beta}
\delta_{m,m'}
\label{eq133hg}
\end{eqnarray}
where $\alpha=\pm 1/2$ denotes the spin and $m, m'=1,...,8$ denotes the 
sites of the internal crystal.
The appropriate Hamiltonian for the free i-spin system 
is given by \cite{roman}
\begin{eqnarray} 
H_0=- \sum_{m,m'} t(m,m')[c_{\alpha}^{\dagger} (m)
c_{\alpha} (m') +h.c.] 
\label{eq2133hg}
\end{eqnarray}
where the sum is over all crystal sites $m \neq m'$
and t is the tunneling rate between the spins.

When it comes to interactions a suitable framework for discussion 
is given by Heisenberg spin models. These have been 
considered in statistical and solid state physics for a 
long time \cite{spinm1,spinm2,borov}, and they have been used to describe 
magnetic phase transitions and excitations as well as many other phenomena. 
The basic variables are spin vectors $\mathbf S$ defined on each 
site of the internal crystal.
They are related to the original creation and 
annihilation operators eq. (\ref{eq133hg}) via 
\begin{eqnarray}
\mathbf{S}(m) = \frac{1}{2} c_{\alpha}^{\dagger} (m) 
\boldsymbol{\tau} _{\alpha\beta} c_{\beta} (m)
\label{eq557}
\end{eqnarray}
where $\boldsymbol{\tau}$ is the triplet of internal Pauli matrices. 

What are the symmetries of this system? 
There are 2 continuous symmetries, associated 
with the conservation of internal charge 
and spin, respectively. Namely, the 
Hamiltonian $H_0$ is invariant under
U(1) transformations 
$c_{\alpha} (m) \rightarrow \exp(i\theta_0)c_{\alpha} (m)$ 
for arbitrary (constant) angle $\theta_0$ 
and under (constant) SU(2) transformations 
$c_{\alpha}(m)\rightarrow \exp(i\boldsymbol{\theta\tau})c_{\alpha}(m)$.
The point is that without interaction the i-spins are fixed to the 
crystal sites but otherwise can rotate freely 
so that the system shows an overall internal SU(2)
symmetry. When the interaction is switched on, the 
SU(2)-breaking mignetic ground state will be formed and the mignons 
will appear as spin vector fluctuations around the ground state. 


There are also several discrete symmetries. 
First of all, there is the point group symmetry dictated by the 
discrete nature of the internal crystal. 
To be specific 
a tetrahedral arrangement of spins 
is chosen with point group symmetry $S_4$ 
and 8 spin vectors forming an internal 'mignetic 
molecule' \cite{magmol,schmi}. Counting the degrees of freedom 
one easily sees that there are 3$\times$8=24 
quasi-particle states to be expected, which can be ordered 
according to the symmetry of the (mignetic) ground state, 
cf. eq. (\ref{eq833hg}) below.
Note that $S_4$ is not 
a chiral symmetry because it contains improper rotations 
in the form of reflections of a plane. 
However, the spin vector is a pseudovector, 
and this implies that the combined point and 
mignetic transformations can form a chiral group - the 
Shubnikov group to be introduced later.
Finally, the Hamiltonian is real ($H_0=H_0^\ast$), 
a signature of time reversal invariance.
Just as SU(2) and the reflection symmetries, the time reversal invariance 
will be broken by the mignetic ground state. 
At this point the existence of an internal time variable s 
which describes processes in the internal 
crystal and differs from physical time t, is mandatory, 
not only because it naturally leads to 
internal spin waves of antifermion spins 
which can be used to describe the quantum numbers of 
antiquarks and antileptons but also because 
the breaking of internal time reversal invariance will play an important 
role for the discussion of the ordered mignetic structures 
presented below. 
That is the reason why I started with SO(6,2) as the complete
symmetry of the whole space before the compactification - instead of
SO(6,1) as was done in ref. \cite{lamm}. 

\begin{figure}
\begin{center}
\epsfig{file=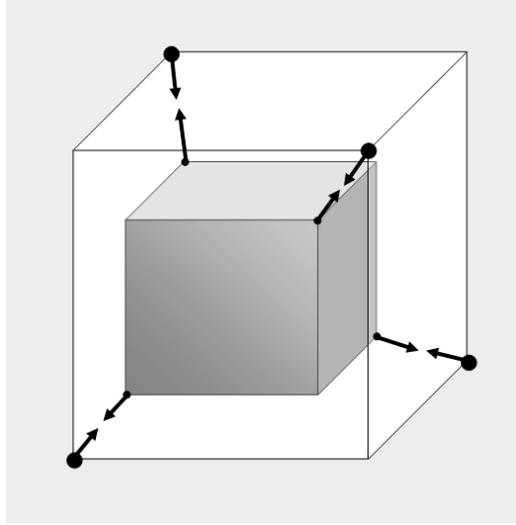,height=7cm}
\bigskip
\caption{Mignetic ground state with 8 spin vectors arranged as follows: 
the corner points of the outer tetrahedron (big black dots) are given by 
the coordinate vectors 
$(1,1,1)$, $(1,-1,-1)$, $(-1,1,-1)$ and $(-1,-1,1)$ and the 
spin vectors are chosen as to point to its centre $(0,0,0)$, 
i.e. the spin vectors sitting on the 4 sites are the 
negative of the coordinate vectors. 
The remaining 4 points lie on the inner tetrahedron (small black dots) 
which is obtained from the outer by multiplying the above coordinates  
by a common shrinking factor $<1$, and the spinvectors of the inner 
tetrahedron are oriented opposite to those of the first one.
The tetrahedra themselves have the tetrahedral group $S_4$ as 
point group symmetry. 
From the pseudovector property of the spin vectors it can 
be shown that the spin system has $A_4 + S ( S_4 - A_4)$ symmetry.}
\nonumber
\end{center}
\end{figure}

\begin{figure}
\begin{center}
\epsfig{file=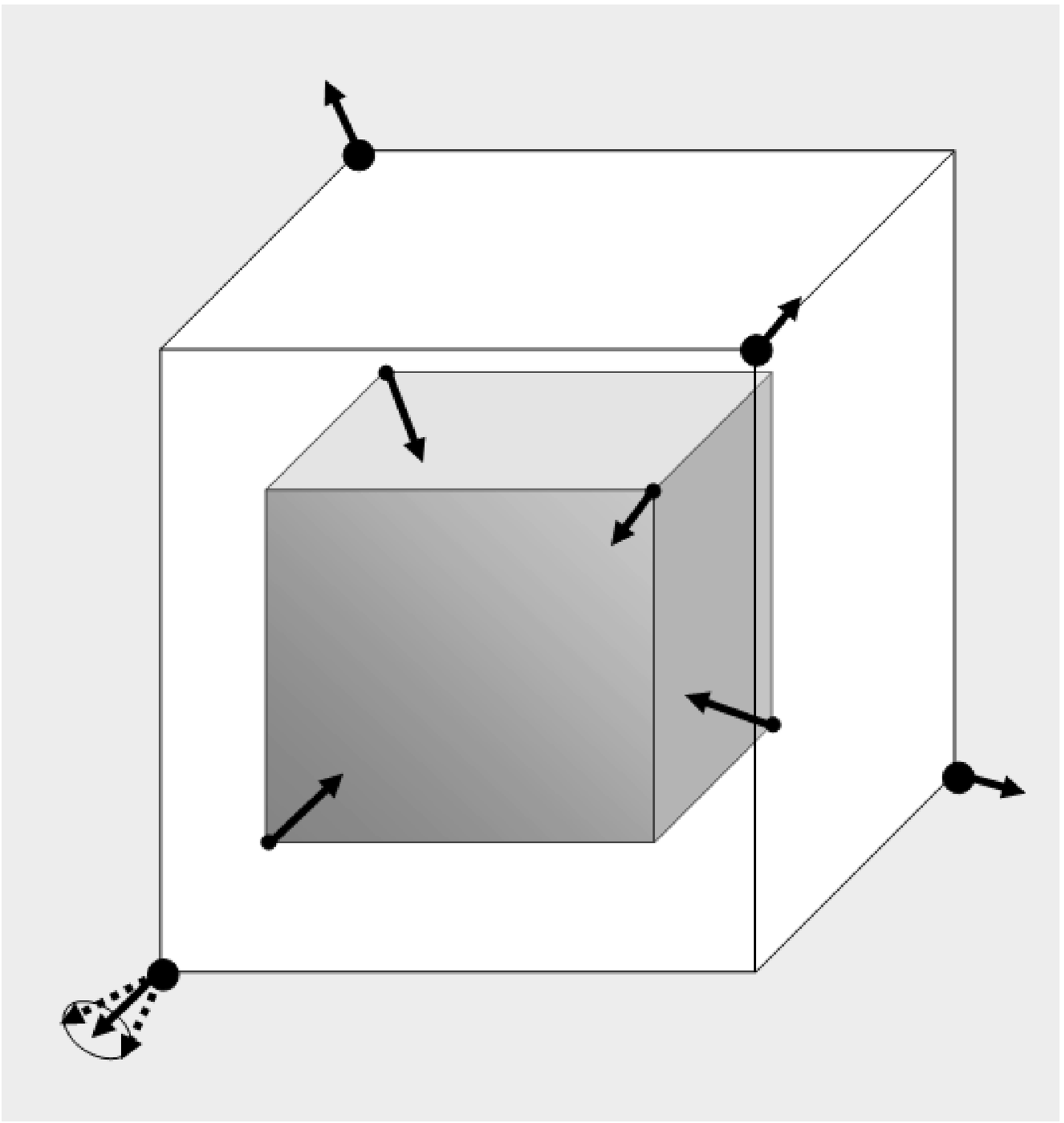,height=7cm}
\bigskip
\caption{A ground state similar to fig. 1, however 
with opposite chirality. As compared to fig. 1 all spins 
point in opposite directions. Also shown is the behavior 
of one of the spin vectors in an excited state. Such 
an excitation is obviously not identical to its chiral 
counterpart, i.e. to the states derived from fig. 1}
\nonumber
\end{center}
\end{figure}

It is well known that one should use Shubnikov
groups (which are sometimes called black-and-white groups) \cite{shub,proj} 
instead of ordinary point groups to classify the 
spectrum of spin wave excitations. More concretely, we shall 
use the Shubnikov group $A_4 + S ( S_4 - A_4)$ instead of the
pyritohedral group $A_4 \times Z_2$ considered in the article \cite{lamm}. 
Here $S_4$ is the symmetry group of a regular tetrahedron, and $A_4$ 
its subgroup of proper rotations (i.e. without 
reflections).\footnote{These groups have been discussed in connection with 
neutrino and family mixing by many authors. For a review see \cite{altarelli}.}
S denotes the (internal) time inversion operation, which in an elegant 
way replaces the rather unnatural $Z_2$-factor in $A_4 \times Z_2$.
Instead of eq. (4) of the paper \cite{lamm}, the 24 mignon 
states are then given by 
\begin{eqnarray} 
A(\nu_{e})+A'(\nu_{\mu}) +A''(\nu_{\tau}) +T(d)+T(s)+T(b)+ \nonumber \\
A_s(e)+A'_s(\mu)+A''_s(\tau) + T_s(u)+T_s(c)+T_s(t) 
\label{eq833hg}
\end{eqnarray}
where $A$, $A'$, $A''$ and $T$ are singlet and triplet 
representations of $A_4$ 
and the index s denotes
genuine representations of the Shubnikov group $A_4 + S ( S_4 - A_4)$
\cite{shub,bata,borov}. 
The 24 d.o.f. in eq. (\ref{eq833hg}) 
correspond in fact to 8 spins each with 3 possible 
directions of the spin vector, as predicted above, 
and the multiplet structure in eq. (\ref{eq833hg}) 
can be obtained by the methods 
described for example on the Bilbao crystallographic server\cite{bilbao,bilbao1}.

Note that this structure is in fact quite unique. 
None of the other possible Shubnikov groups \cite{shub} show a 
pattern of the type eq. (\ref{eq833hg}). 
Most of them do not even have triplets, and if so, one usually 
finds doublets as well. 

To understand the physics it is useful trying to construct a
ground state, for which the 24 states eq. (\ref{eq833hg}) 
represent (internal spinwave) excitations. In other words, one is looking for 
a static system with 8 internal spins and symmetry group $A_4 + S ( S_4 -
A_4)$. The simplest 'mignetic molecule'  \cite{magmol,schmi} 
of this kind consists of two
regular tetrahedra with spin vectors arranged as in figure 1. Note that 
this ground state shows a rather strong 
type of antiferromagnetic order. Firstly, the spins in
each tetrahedron add up to zero. Furthermore the spins appear in pairs
with partners coming from both tetrahedra and which are oriented
oppositely.

As a consequence, the 
(internal) Heisenberg spin SU(2) symmetry of the mignetic 
system is broken
as well as the internal point symmetry $S_4$: 
\begin{eqnarray} 
SU(2) \times S \times S_4 \rightarrow A_4 + S (S_4-A_4)
\label{eq83311hg}
\end{eqnarray}
The remaining Shubnikov symmetry $A_4+ S ( S_4
- A_4)$ does not contain any reflections, because improper 
rotations $S_4-A_4$
appear only in combination with the time reversal operation S.
Therefore it is a chiral group (just as $A_4 \times Z_2$), a property
which was essential in the paper \cite{lamm} 
to derive the parity violation of the
weak interactions.

Note further that internal time reversal S is itself
broken, as can be seen easily, because it is {\it not} an element of
the Shubnikov group $A_4 + S ( S_4 - A_4)$. 
Only combinations of the form $SR$, where $R \in S_4 - A_4$ is an 
improper rotation, are symmetries of the system. 
The point is that applying S (or R) to the ground state fig. 1 
one will obtain a different state (fig. 2) with higher energy and 
opposite chirality, whose excitations have nothing to do with the 
excitations of fig. 1.

All states in eq. (\ref{eq833hg}) are therefore chiral states 
(with respect to the internal chiral structure), 
the difference between $A$ and $A_s$, $A'$ and $A_s'$ etc excitations 
being mainly odd and even behavior under transformations $SR$. 
Since they are different multiplets, they will in 
general have different masses. On the level of quarks and
leptons this gives different masses to weak isospin partners.
 
How is this linked to the breaking of the Standard Model gauge group
SU(2)$_L$? My conjecture is that the fundamental spinor 
F introduced in the r.h.s. of eq. (\ref{eq3uijj}) can be used 
to form a vacuum condensate $\langle\bar F_R^+ F_L^+\rangle\neq 0$. 
(The lower index R or L denotes chirality in Minkowski space 
and the upper index + or - denotes the spin direction in 
internal space.)
As discussed in section 6 of ref.\cite{lamm}, internal and external 
chirality can be related via octonion multiplication. 
Therefore, as soon as the discrete internal chiral structure 
fig. 1 is formed, right handed fermions are excluded from the 
weak interactions. Accordingly, the right handed components of 
the fundamental fermion F effectively 
turn out to be singlets of the Heisenberg SU(2) 
while the left handed remain as doublets. 
Accordingly, an effective Higgs doublet 
\begin{eqnarray} 
H \propto \bar F_R^+
\left( \begin{array}{c}
F_L^+\\
F_L^-
\end{array}\right)
\label{eq3hh866}
\end{eqnarray} 
can be constructed to mimic the Standard Model Higgs field, 
and a symmetry breaking quite 
similar to NJL or technicolor models takes place.
More details will be given in a future publication.

In conclusion I have tried to analyze the spectrum of quarks and leptons
on the basis of dynamics taking place in a 3-dimensional internal space. 
Since this is a rather unusual approach 
it may seem hard to understand where it comes from and to where it will lead,
in particular because I have mostly restricted myself to the internal processes, 
and did not consider the propagation and interactions 
of the mignons in Minkowski space. 
Therefore I would like to include a comment how the model can be extended 
to obtain a more comprehensive scenario. 

In fact there is a certain physical picture in my mind 
where our universe resembles a huge crystal of 
molecules, each 'molecule' with 8 'atoms' and each 
atom with a spin degree of freedom, which can be 
excited to form {\it intra}-molecular excitations. 
Such excitations are known to exist in real 
molecular crystals \cite{vibron1} and can propagate 
through these crystals to become {\it inter}-molecular 
quasi-particles. The way this happens is by transfering the 
excitation from one molecule to the other. 

The main difference with mignons is that these have higher 
dimensionality, i.e. the 'molecules' extend to internal dimensions 
which are orthogonal to physical space. 
Still it is possible that mignons in an internal space $I_x$
over a given spacetime point $x$ are able to excite mignons in 
neighbouring internal spaces $I_y$ and thus can travel 
as quasi-particles through  
Minkowski space with a certain wave vector $\vec k$ which is to be 
interpreted as the physical momentum of the quark or lepton. 

Furthermore there is the possibility of interactions 
when two such 'quasi-particles' collide. I have not yet analyzed this in detail, 
but as argued in section 4 of ref. \cite{lamm} it is conceivable that a gauge 
structure of the interactions may arise. 



\end{document}